\documentstyle[version2,aps,preprint]{revtex}
\begin{document}
\draft
\begin{title}
\begin{center}
Observation of the Holstein shift in high $T_c$ superconductors \\
with thermal modulation reflectometry
\end{center}
\end{title}
\author{O. V. Dolgov}
 \begin{instit}
	P. N. Lebedev Physical Institute, Russian Academy of Science \\
        Moscow 117924, Russia.
 \end{instit}
\moreauthors{A.E. Karakozov, A.A. Mikhailovsky}
 \begin{instit}
	L. F. Vereschagin Institute for High Pressure Physics\\
	Troitsk, Russia
 \end{instit}
\moreauthors{B.J. Feenstra, and D. van der Marel}
 \begin{instit}
        Materials Science Center, Solid State Physics Laboratory\\
        University of Groningen, Nijenborgh 4\\
        9747 AG Groningen, The Netherlands
 \end{instit}
\begin{abstract}
We use the experimental technique of thermal modulation
reflectometry to study the relatively small temperature dependence
of the optical conductivity of superconductors.
Due to a large cancellation of
systematic errors, this technique is shown to a be very sensitive probe of
small changes in reflectivity.
We analyze thermal modulation reflection spectra of single
crystals and epitaxially grown thin films of YBa$_2$Cu$_3$O$_{7-\delta}$
and obtain the ${\alpha_tr}^2F(\omega)$ function in the normal state,
as well as the superconductivity induced changes in reflectivity. We present
detailed model calculations, based on the Eliashberg-Migdal
extension of the BCS model, which show good qualitative and quantitative
agreement with the experimental spectra.
\\
{\em Materials Science Center, Internal Report no VSGD93.12.th}\\
Physica C, in print
\end{abstract}
\newpage
\section{Introduction}
Optical spectroscopy in the infrared region provides a valuable tool to study
low-energy excitations in high $T_c$ superconductors. In principle such
measurements can provide information about the superconducting
energy gap (if any) and the pairing mechanism.
A lot of discussion is still focussed on the
problem of how to distinguish electronic contributions from those
which are due to phonons,
and to what extent these two are coupled. The latter deserves special
attention, because in the conventional superconductors
electron-phonon coupling is believed to be responsible for superconductivity.
While the role of electron-phonon coupling is not very clear
in the new high $T_c$ materials,
there exists a tendency in the theoretical community to consider conventional
coupling mechanisms such as electron-phonon as a necessary ingredient to
obtain pairing. It is not yet clear however whether an exceptionally large
electron-phonon coupling is really required, or whether a small germ suffices,
which is then further 'boosted' by other mechanisms such as interlayer
pairing (as was recently proposed by P.W. Anderson\cite{phil}).
This uncertainty emphasizes the importance of a proper understanding of
electron-phonon coupling in these materials if we want to make progress
towards a theory of high $T_c$ which is of practical importance, e.g.
able to predict $T_c$ for new compounds. Although much data of high
quality has been obtained on a wide
variety of cuprate high $T_c$ superconductors,
it has turned out difficult to obtain a clear interpretation of these data. In
particular there have been many papers where
the observation of a 'clean' gap has been claimed,
but closer inspection reveals, that there is a considerable amount of
residual absorptivity at energies below the presumed gap.
Some of the difficulties are connected with experimental
problems concerning the complicated structure of such systems,
but others are related to
the theoretical interpretation of the results obtained. Two examples are that
the value these gaps would have is far too high to be compatible with a
simple BCS picture, and the position of these features doesn't appear
to shift to zero frequency if $T \rightarrow T_c$. Instead
a gradual filling of the gap region is observed with a
change of slope of the intensity in this region
at the phase transiton, which can be fitted in a
phenomenological way using a two-fluid
description of the superconducting state.

In the conventional theory superconductivity
is a consequence of electron pairing
mediated by the exchange of the Bose-like excitations. This pairing leads
to the appearance of an energy gap and hence to zero
absorption for frequencies less than
$2\Delta$. In the optical data it is possible
to determine this threshold by measuring the
departure of the reflectivity from unity, or
of the absorptivity from zero. The most
informative value is the optical conductivity
($\sigma$). But in order to calculate $\sigma$ it
is necessary to determine with great accuracy
small departures from $100 \%$ reflectivity,
followed by a Kramers-Kronig analysis to obtain the phase of the
reflectivity. When the absolute reflectivity is close to $100 \%$
the progression of experimental errors in this procedure often leads to
large uncertainties in the conductivity.

For experimental reasons 2$\Delta$ was determined
in conventional superconductors as a
maximum of a ratio of the reflectivity in the SC state
to the normal state. For HTS there are
several difficulties when one tries to apply
this procedure due the high critical
temperature $T_c$ itself. Between zero
temperature and $T_c$ some intrinsic changes
in the electronic subsystem occur, due to
which we have practically two different substances
at low (helium) and normal state temperatures. This
is more important if the mechanism
of pairing is of an electronic nature.

\section{Thermal Modulation Reflectometry}
Another method of the determination of the energy gap has been
used by Abel {\em et al.} \cite{abel}.  It provides
a possible way to avoid some of the afore mentioned difficulties
(e.g. problems with the absolute
value of reflectivity and Kramers-Kronig transformations).
Abel {\em et al.} determined the ratio of optical
reflectivities at two close temperatures
$R(\omega , T)/R(\omega , T + \delta T)$,
and the maximum of this ratio was ascribed to the
superconducting energy gap $ 2 \Delta
(T) $. This feature reflects the fact that the main change
below $T_c$ is due to the decrease of $\Delta$ in the
single-particle excitation spectrum.
However, in real superconductors the situation is more complicated.
First, for $T>0$ the gap itself is not well
defined. Second, we have a strong temperature dependence of the number
of thermally excited Bose-like quasi-particles.
As we shall see below, this is the dominating effect, and
gives us a possibility to extract
information about the spectrum of these intermediate bosons.
The third complication is connected to the so-called
Holstein shift. The optical conductivity
contains contributions from the intermediate bosons which are responsible
for the pairing {\em because} they are coupled
to the electron-hole excitations
(boson-assisted conductivity, see e.g.  \cite{holstein,pballen}).
The difference between the normal and the superconducting state is
the following: A threshold point of an absorption
which in the former case takes place at
the characteristic boson frequency $ \Omega_0 $
should in the latter case be shifted to
$ \Omega_0 + 2 \Delta $. This effect was observed in
conventional superconductors and played an important
role in establishing the phonon-mediated
nature of superconductivity in these systems  \cite{joyce,farnworth}.
However, the absence of such a shift in
the reflectivity spectra of high $T_c$ superconductors
is one of the most important objections
against the conventional mechanism of superconductivity in these compounds
\cite{tanner+timusk,zack90}.
The thermo-modulation method, in which the
ratio of reflectivities at two close
temperatures is analyzed \cite{abel}, gives the possibility
to observe the boson anomalies in the
normal as well as in the superconducting state. In the present paper
we make a detailed comparison between model calculations based on
strong-coupling theory, and experimental thermo-modulation reflection
spectra. The data used in the present analysise are in good agreement with
those obtained by Abel {\em et al.}, but cover a wider range of
frequencies and temperatures. \\
The main advantage of measurements of $R(\omega,T)/R(\omega,T+\delta T)$
between two close temperatures is
that all thermal effects connected with extrinsic factors
(e.g. the experimental setup, spurious signals)
are compensated and the ratio reflects only the temperature
dependence of the occupation numbers of the bosonic and fermionic excitations.
If the change of temperature $\delta T \ll T$, it is possible to expand
\begin{equation}
R(\omega,T)/R(\omega,T+\delta T) = 1+r(\omega,T)\delta T
\end{equation}
 From the expression for the reflectivity at normal incidence
$R=|(1-\sqrt{\epsilon})/(1+\sqrt{\epsilon})|^2$ one obtains the
following exact expression for the thermo-reflectance coefficient
\begin{equation}
        r(\omega, T)
        \equiv - \partial \ln R(\omega,T) / \partial T
        =  - 2 Re \left[\frac{\partial \epsilon / \partial T}
                           {(\epsilon-1)\sqrt{\epsilon}} \right]
\label{eq:r-exact}
\end{equation}
Although in the comparison which we will make to
experimental optical data we will use numerical
calculations based on the full solution of
the Eliashberg equations, we will make some further approximations in order
to demonstrate that $r(\omega, T)$ reflects the $\alpha_{tr}^2F$ function.
We first notice, that for a good metal in the frequency region under
consideration  ($\omega \ll \omega_p$) $|\epsilon'| \gg 1$, so that
we may write $r(\omega, T)=4 Re \partial(\epsilon^{-1/2})/\partial T$.
As can be seen from Fig. 8 of Ref. \cite{marel}
$0.25<\epsilon''/\epsilon'<0.4$ for the relevant frequency range in the
superconducting state. Although in the next section we will make
a comparison between experiment and theory using the exact
expression for the thermo-reflection (Eq. 2),
in the present qualitative discussion we only consider the 'clean' limit,
where $(\epsilon''/\epsilon')=(\omega\tau)^{-1}$ may be treated
as a small parameter in a Taylor series expansion. The leading order
of this expansion is
\begin{equation}
       r(\omega, T)
       \approx 2\frac{\partial}{\partial T}
       \frac{\epsilon''}{|\epsilon'|^{3/2}}
\end{equation}
For the description of the dielectric function in the
normal state we use the extended Drude expression\cite{jim}
\begin{equation}
 \epsilon(\omega,T) = \epsilon_{\infty} -
 \frac{\omega_p^2/\omega^2}
      {m^*(\omega)/m+i/(\omega\tau(\omega))}
\end{equation}
%
We will make the approximation here,
that the main temperature dependence
enters through the parameter $\tau(\omega)$.
As we consider here the range of frequencies
where $\omega\tau \gg 1$, we obtain the simple expression
\begin{equation}
 r(\omega,T)=\frac{2}{\omega_p}\frac{\partial\tau^{-1}}{\partial T}
\end{equation}
As is shown in the Appendix, for sufficiently
low frequencies the temperature derivative of the
optical scattering rate is proportional to the
transport spectral function, whereas at high frequencies it
is proportional to $\lambda_{tr}$, so that
\begin{equation}
r(\omega,T) =
 \frac{4\pi}{\omega_p}\left\{
 \begin{array}{ll}
 \lambda _{tr}
 & (T \gg \Omega_0) \\
 &                  \\
 \frac{2\pi^2T}{3\omega}
 \alpha^2_{tr}(\omega)F(\omega)
 & (T \ll \Omega_0) \\
 \end{array}
 \right\}
\label{eq:rnormal}
\end{equation}
So we see, that thermal modulation reflectometry can provide rather
direct experimental information on the transport spectral function.
\\
In the superconducting state it is possible to obtain a useful expression
in the region where
$-\epsilon'\simeq\frac {c^2}{\omega^2\lambda_L^2} \gg
\frac{4\pi\sigma_1(\omega,T)}{\omega}$ (i.e., when $R \simeq 1$),
where $\sigma_1=\omega\epsilon''/(4\pi)$ is the real part
of the optical conductivity.
Here $\lambda_L$ is the London penetration depth. In this case
\begin{equation}
r(\omega,T)=8\pi c^{-3}\lambda^3_{L}\omega^2
            \frac{\partial\sigma_1(\omega,T)}{\partial T}
\label{eq:r2}
\end{equation}
In the Appendix we show that in the superconducting state
the dominant term in $\frac{\partial\sigma_1(\omega,T)}{\partial T}$
is proportional to $\alpha^2_{tr}F(\omega-2\Delta)$.
As a result we obtain, disregarding the slowly varying term
\begin{equation}
 r(\omega,T) =  \frac{8 \pi^3 \omega_p^2\lambda_L^3 T}{3c^3}
                \alpha^2_{tr}(\omega-2\Delta)F(\omega-2\Delta)
\end{equation}
At intermediate temperatures the dielectric function can sometimes
be approximated with the two-fluid interpolation formula
$\epsilon(\omega ,T)=\epsilon_s f_s(T) + \epsilon_n (1-f_s(T))$,
as was supported experimentally \cite{marel} for
YBa$_2$Cu$_3$O$_{7-\delta}$ and theoretically
in the case of strong or intermediate coupling \cite{mikhailovskii},
where $\epsilon_s(\omega)$ and $\epsilon_n(\omega)$ are the dielectric
functions at $T=0$, and at $T\ge T_c$ respectively. The
function $f_s(T)$ is proportional to the number of superconducting
electrons and is assumed to be of the form
$f_s(T)=1-\left( \frac{T}{T_c}\right)^{\nu}$
where $\nu$ is some exponent (usually $\nu=4$).
For $T<T_c$ we see that $r(\omega ,T) \propto
Re[\sigma_n(\omega)-\sigma_s(\omega)]$ and has its maximal
amplitude directly below $T_c$.
\section{Comparison of experimental data with strong-coupling calculations}
In Fig. 1a the experimental values of the thermo reflectance coefficient
$r(\omega,T)$ are displayed
for an ab-oriented thin film of  $YBa_2Cu_3O_{7-\delta}$
with $T_c = 90 K$ \cite{marel} for
temperatures between 30K and 140K.
The ratios were calculated from data differing by 10 K. We see a
feature in the frequency region $\omega \leq 1000 cm^{-1}$
both in the normal state
and in the superconducting state.
We made a similar analysis of the data obtained by Bauer on
a single crystal of YBa$_2$Cu$_3$O$_{6.8}$
with a $T_c$ of 81-86 K \cite{bauer}. As only
reflectivity data at 10, 60 100, and 150K were taken, the ratio's
$\frac{2}{T_2-T_1} \frac{R(T_1)-R(T_2)}{R(T_1)+R(T_2)}$
(displayed in Fig. 1b) are calculated for the
pairs $(T_1,T_2)=$ $(10,60)$, $(60,100)$
and $(100,150)$. Although one expects deviations from
a pure thermal derivative in this case, the thermal modulation
spectra obtained from both sets of data are actually quite similar.
The origin of the larger value of $r$ between 100 K and 150 K in Bauer's
data is not clear, and may be due to the lower oxygen-concentration
with a $T_c$ of 80-85 K in this sample.
In the normal state all features end at the
frequencies near $700 cm^{-1}$, which
corresponds to the range of phonon frequencies for these materials.
According to expression 6
the shape of $r(\omega ,T)$ should be
proportional to the spectrum of intermediate phonons
$\alpha^2_{tr} F (\omega)$.
This gives us the possibility to speak
about the intermediate boson contribution to the optical
properties of high $T_c$ systems.

Let us first calculate the thermo-modulation spectra using the weak coupling
BCS model in the clean limit, either
assuming a single gap at $220 cm^{-1}$ (Fig. 2a) or a distribution
of gaps between $0$ and $500 cm^{-1}$ (Fig. 2b).
We see, that the calculated $r (\omega ,T)$ has only
one feature corresponding to the energy gap.
Also the calculated thermomodulation effect on the
reflectivity is much larger
than the experimental values.

To obtain a better understanding of these
features we carried out model calculations of the
reflectivity assuming an $\alpha^2F$ function with a single broad
peak at $\omega_0 = 350 cm^{-1}$,
the constant of interaction being $\lambda=1.5$,
and the bare $\omega_{p} = 3 eV$, which gives a critical
temperature $T_c = 87K$. This shape
of $\alpha^2 F(\omega)$ was used in  \cite{karakozov}
These parameters lead to a linear dependence of the resistivity
in the normal state, with a slope that
corresponds to experimental values \cite{igor}.
To match the 164K data it is either necessary to
take or a larger $\omega_0$ or larger $\lambda$ ($\approx 2.5$).
The results are shown in Fig. 3.

According to expression 8, for $T \ll T_c$ the spectrum
of the intermediate bosons should be shifted by $2\Delta$.
The experimental data indeed
show such a shift and make it possible
to estimate the value of $2\Delta$ (if $\Delta_k$ is
actually a distribution of gap-values
due to {\em e.g.} anisotropic pairing, the shift corresponds
to a gap-value averaged over the Fermi-surface) to be
$250$ to $300$ (cm$^{-1}$), corresponding to a ratio
$2 \Delta /T_c \approx 4$ to $5$. It is interesting
to note the negative contribution to $r(\omega,T)$
above $2 \Delta + \Omega$, which is a consequence of the
modification of the optical conductivity
(second term in Eq.\ref{eq:final}). At intermediate
temperatures the ratio $r(\omega, T)$ behaves
according to the two-fluid model and reaches
a maximal amplitude just below $T_c$.

\section{Conclusions}
We demonstrate that thermal modulation
reflectometry can be used to record small changes in reflectivity
of superconductors, even if the reflectivity
itself is close to $100 \%$. We observe that the thermal
modulations of superconducting $YBa_2Cu_3O_{7-\delta}$
are quite small, but well reproducible from sample
to sample. Although the observed effects are
much smaller than a weak coupling BCS-type calculation in the clean limit,
qualitatively good agreement with a strong-coupling type
model calculation is obtained, where $\lambda=1.5$,
and $\omega_{p} = 3 eV$ is
assumed, and an $\alpha^2F$ function with a single broad
peak at 350 (cm$^{-1}$) is taken.
\section{Acknowledgements}
This work was completed with financial aid
from the Nederlandse Organisatie voor Wetenschappelijk Onderzoek. We
thank Prof. Dr. L. Genzel, and Dr. M. Bauer for making their data-files
available in digital form.
\section{Appendix}
\subsection{Normal state}
Let us first consider the reflectance in the normal state.
The temperature dependence of the optical scattering rate can be expressed
using the relation of Ref.\cite{shulga} (valid for $\omega\tau \gg 1$)
\begin{equation}
\begin{array}{l}
 \tau(\omega ,T)^{-1} =  \\
 \frac {\pi}{\omega}\int^{\infty}_{0}d\Omega\alpha_{tr}^2(\Omega)F(\Omega)
 \left[ 2\omega \coth \left(\frac{\Omega}{2T}\right)
 + (\omega-\Omega) \coth \left(\frac{\omega-\Omega}{2T}\right)
 - (\omega+\Omega) \coth \left(\frac{\omega+\Omega}{2T}\right) \right]
\end{array}
\end{equation}
where $\alpha^2_{tr}(\omega) F(\omega)$ is a transport spectral
function of the electron-boson interaction
\begin{equation}
\alpha^2_{tr}(\omega) F(\omega)=\frac{N(0)}{4v_F}
\ll \mid M_{kk'}
 \mid^2 (\vec{V}_k - \vec{V}_{k'})^2
 \delta (\Omega_{\vec{k}-\vec{k}'} - \omega ) \gg
\end{equation}
where $\Omega$ and $M$ are an intermediate boson frequency
and a matrix element, and $\ll .... \gg$
denotes an average  over  the Fermi surface.
In the limiting cases, where the temperatures are either much smaller
or much larger than the characteristic frequency $\Omega_0$, we have
\begin{equation}
 \tau(\omega,T)^{-1} =
 \left\{
 \begin{array}{ll}
 4\pi T\int^{\infty}_{0}
 d(\ln\Omega)\alpha^2_{tr}(\Omega)F(\Omega)=2\pi\lambda _{tr}T
 & (T \gg \Omega_0) \\
 &                  \\
 \frac {2 \pi}{\omega}\int^{\omega}_{0}d\Omega
 \alpha^2_{tr}(\Omega)F(\Omega)(\omega-\Omega)+\frac{2\pi^3}{3\omega}T^2
 \alpha^2_{tr}(\omega)F(\omega)
 & (T \ll \Omega_0) \\
 \end{array}
 \right\}
\label{eq:tau}
\end{equation}
The first term has been previously obtained by Allen  \cite{pballen}.
The first derivative with respect to temperature becomes
\begin{equation}
\frac{\partial\tau(\omega,T)^{-1}}{\partial T} =
 \left\{
 \begin{array}{ll}
 2\pi\lambda _{tr}
 & (T \gg \Omega_0) \\
 &                  \\
 \frac{4\pi^3T}{3\omega}
 \alpha^2_{tr}(\omega)F(\omega)
 & (T \ll \Omega_0) \\
 \end{array}
 \right\}
\label{eq:dratedt}
\end{equation}
\subsection{Superconducting state}
For $T \ll T_c$ the expression for the boson-assisted
conductivity was obtained by
Allen \cite{pballen}
\begin{equation}
\begin{array}{l}
\sigma_1(\omega ,T) = \\
 \frac {\pi e^2}{6 \omega^2} \ll \mid M_{kk'}
 \mid^2 (\vec{V}_k - \vec{V}_{k'})^2 f_k (1 - f_{k'})
 \left( 1 - \frac {\epsilon_k \epsilon_{k'} -
 \Delta_k \Delta_{k'}}{E_k E_{k'}} \right)
 \delta (E_k + E_{k'} + \Omega_{\vec{k}-\vec{k}'} - \omega ) \gg
\end{array}
\end{equation}
Here $E_k = \sqrt{\epsilon_k^2 + \Delta_k^2} $
is a quasiparticle spectrum, $\Delta_k$
is the superconducting gap and $f_k$ is the Fermi-distribution.
For low temperatures $T \ll
\Delta$ it is possible to neglect the temperature dependence of the gap, and
we obtain
\begin{equation}
\begin{array}{l}
 \frac{\partial \sigma_1(\omega,T)}{\partial T} =
 \left(\frac{\pi e^2}{6\omega^2}\frac{2\pi^2 n T}{m}\right)   \\
   \left\{
     - 4\alpha^2_{tr}F(\omega-2\Delta)
     + \int_{0}^{\omega-2\Delta}d\Omega\alpha_{tr}^2F(\Omega)
       \frac{\sqrt{(\omega-\Omega)(\omega-\Omega-2\Delta)}}
       {\Delta(\omega-\Omega-\Delta)}
   \right\}
\end{array}
\label{eq:final}
\end{equation}
The first term is proportional to the Eliashberg function
$\alpha^2_{tr}(\omega) F(\omega)$ shifted by $2\Delta$ (Holstein shift).
The second term is weakly varying
with frequency and is a consequence of a
modification of the conductivity at $\omega > 2\Delta+\Omega$.
\pagebreak

\figure{a: Experimental $r(\omega,T)$ with $\vec{E} \perp \vec{c}$
for an epitaxial thin film of
YBa$_2$Cu$_3$O$_{7-\delta}$. From top to bottom:
T= 35, 55, 75, 95, 115 and 135 K. The curves have been shifted vertically
with 0, -0.0005, -0.0010 K$^{-1}$ {\em etc}. \\
b: The same for a single crystal. Temperatures are 35, 80 and 125 K. Vertical
offsets have been given of 0, -0.001 and -0.002 K$^{-1}$. }

\figure{a: Theoretical calculation of $r(\omega,T)$ based on the BCS model
($2\Delta=3.5 k_BT_c$) using the method of Ref. \cite{rainerb}. Temperatures
are 35, 55 and 75 K from top to bottom. Vertical offsets have
been given of 0, -0.001, and -0.002 K$^{-1}$. \\
b: Theoretical calculation of $r(\omega,T)$ based on a BCS-like model with
a gap distributed between 0 and $8 k_BT_c$. Temperatures and offsets are as
in Fig. 2a.}

\figure{Theoretical calculation of $r(\omega,T)$ based on the strong coupling
formalism with parameters as explained in the text.  From top to bottom:
T= 35, 55, 75, 95, 115 and 125 K. The curves have been shifted vertically
with 0, -0.0005, -0.0010 K$^{-1}$ {\em etc}. }

\end{document}